\documentclass{mem}
\usepackage{natbib}\usepackage{txfonts}\usepackage{balance}
\usepackage{flushend}
\usepackage{graphicx}
\usepackage[breaklinks,pdftex]{hyperref}
\idline{75}{282}
\begin{document}
\def\teff{$T\rm_{eff }$}
\def\kms{$\mathrm {km s}^{-1}$}

\title{
Virtual reality for the analysis and visualization of scientific numerical models
}


\author{
S. \,Orlando\inst{1} \and M. \,Miceli\inst{2,1} \and U. \,Lo Cicero\inst{1} \and S. \,Ustamujic\inst{1}
          }

\institute{
INAF-Osservatorio Astronomico di Palermo, Piazza del Parlamento 1, 90134, Palermo, Italy
\email{salvatore.orlando@inaf.it}
\and
Dipartimento di Fisica e Chimica E. Segr\`e, Universit\`a degli Studi di Palermo, Via Archirafi, 36, 90123, Palermo, Italy\\
}

\authorrunning{S. Orlando}

\titlerunning{Virtual reality for analysis and visualization of scientific models}

\date{Received: Day Month Year; Accepted: Day Month Year}

\abstract{The complexity of the data generated by (magneto)-hydrodynamic (HD/MHD) simulations requires advanced tools for their analysis and visualization. The dramatic improvements in virtual reality (VR) technologies have inspired us to seek the long-term goal of creating VR tools for scientific model analysis and visualization that would allow researchers to study and perform data analysis on their models within an immersive environment. Here, we report the results obtained at INAF-Osservatorio Astronomico di Palermo in the development of these tools, which would allow for the exploration of 3D models interactively, resulting in highly detailed analysis that cannot be performed with traditional data visualization and analysis platforms. Additionally, these VR-based tools offer the ability to produce high-impact VR content for efficient audience engagement and awareness.}

\maketitle{}
\section{Introduction}

The development of (magneto)-hydrodynamic (HD/MHD) models of astronomical objects and phenomena can be crucial for our understanding of the structure, dynamics and energetics of these phenomena and for the analysis and interpretation of astronomical observations. These models are described by set of HD/MHD equations which can be solved through sophisticated parallel numerical codes as, for instance, the codes for astrophysical plasmas FLASH (\citealt{2000ApJS..131..273F}) and PLUTO (\citealt{2012ApJS..198....7M}). Running these codes requires high-performance parallel computing systems (using thousands of processors in parallel) and a significant number of numerical resources (in terms of CPU hours and storage memory). The codes are optimized to reduce the computational cost and to improve the parallel efficiency when using several thousands of processors. Typical 3D HD/MHD simulations are executed using of the order of 10 thousands CPUs on parallel supercomputers and require millions of computer hours for the whole computation.

Over the years, the HD/MHD models of astronomical phenomena have become increasingly accurate and complex, also thanks to the improvement of the algorithms for solving the system of HD/MHD equations and to the greater amount of numerical resources available for scientific research. Today, fully three-dimensional (3D) HD/MHD simulations of astrophysical phenomena contain a great wealth of scientific information, which can be difficult to unravel and extract, and produce large amounts of data. For these reasons, modern 3D HD/MHD simulations pose a challenge for their analysis and standard data visualization for scientific purposes.

In the last decade, the potential of virtual reality (VR) hardware and software has started to be exploited in different fields for public outreach and education with excellent results. For example, online digital media stores and education and public outreach websites (such as those promoted by NASA\footnote{\url{https://chandra.si.edu/vr/}} and eduINAF\footnote{\url{https://edu.inaf.it}}) offer, among other things, high-impact VR content in Astrophysics and Space Sciences. Unfortunately, the routine use of VR environments for analyzing and visualizing models for scientific research is still in its infancy.

Driven by the need for efficient tools for analyzing and viewing 3D HD/MHD models, in 2019 we launched 3DMAP-VR (3-Dimensional Modeling of Astrophysical Phenomena in Virtual Reality; see Sect.~\ref{sec2}), a laboratory for the development and testing of VR-based tools for the analysis and visualization of numerical scientific models (\citealt{2019RNAAS...3..176O}) and, more recently, StarBlast (see Sect. \ref{sec3}) a standalone app available for free on Steam that offers a VR tour of the results of stellar explosions accessible even to the general public of non-experts. In 2022 we started the development of a VR-based tool for the analysis of numerical scientific simulations (see Sect. \ref{sec4}). In the next sections, we introduce these projects in more details.

\section{The 3DMAP-VR Project}
\label{sec2}

The project aims to provide an environment for developing tools that exploit VR technologies for visualizing scientific models. In fact, due to the complexity of modern 3D HD/MHD simulations full of details that characterize the structure and evolution of the simulated astronomical objects, untangling the different processes that govern the object under study and extracting relevant information from the model can be a difficult task that cannot be easily resolved by traditional 2D representations of the models as, for instance, in screen displays. For example, the stellar debris ejected after the explosion of a supernova (SN) can be characterized by rather complex structures in which the distributions of the different species of elements intertwine in a chaotic and, apparently, disordered way. In these cases, traditional 2D representations fail to describe inherently 3D structures, making it difficult to identify and follow the evolution of specific features and phenomena of interest.

\begin{figure*}[t!]
\resizebox{\hsize}{!}{\includegraphics[clip=true]{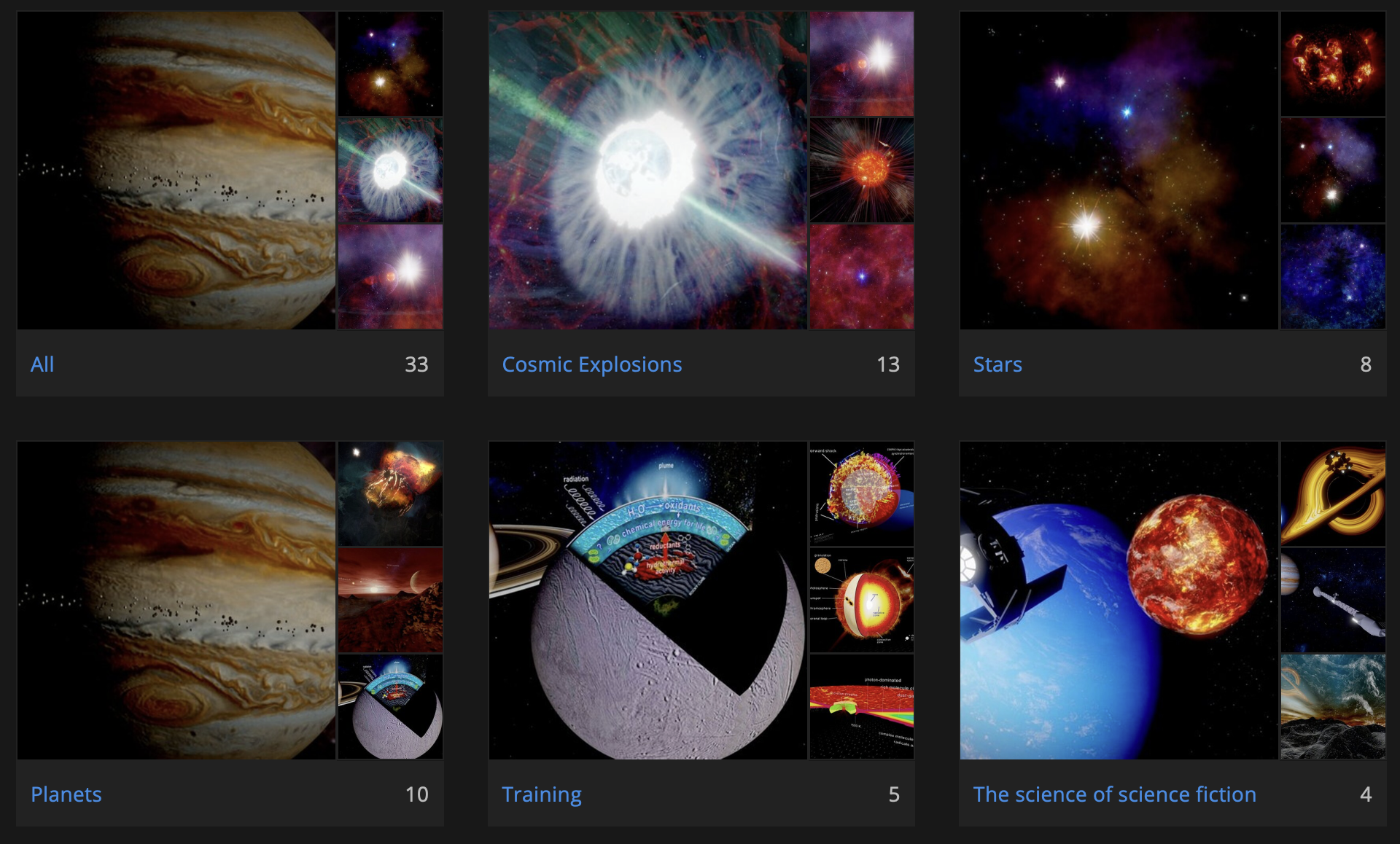}}
\caption{\footnotesize Galleries of models available on Artstation: \url{https://www.artstation.com/saorlando4}.}
\label{fig:artstation}
\end{figure*}

Conversely, exploring models immersively through VR equipment allows scientists to navigate and interact with their MHD models in a more natural and intuitive way. The 3DMAP-VR project aims at providing a VR-based enviromnent for the exploration and visualization of the models. More specifically, in the 3DMAP-VR environment, the workflow for creating VR visualizations of models consists in three steps.

i) {\em Accurate 3D HD/MHD simulations performed for scientific purposes}. First, the 3D HD/MHD models are produced through numerical simulations that are executed on parallel supercomputers. The simulations account for all the relevant physical processes in astrophysical phenomena: gravity, magnetic-field-oriented thermal conduction, energy losses due to radiation, gas viscosity, deviations from proton-electron temperature equilibration, deviations from the ionization equilibrium, cosmic rays acceleration, etc.. 

ii) {\em Tools for model analysis and for meshing and texture production of the different model components}. As a second step, interactive and navigable 3D graphics of the astrophysical simulations are realized by exploiting the tools commonly used by the scientific community for the analysis of data, e.g., Interactive Data Language, YT project, ParaView, Visit, MeshLab, MeshMixer. A mixed technique consisting of multilayer isodensity surfaces with different opacities is used to realize the 3D graphics.

iii) {\em Tools for the creation of objects that can be explored in VR.} Finally, a VR representation of the astrophysical models is realized by uploading the 3D graphics to Sketchfab\footnote{\url{https://sketchfab.com}}, one of the largest open access platforms for publishing and sharing 3D VR and augmented reality content. Once the 3D graphics have been loaded, the VR representations of the model are created through a friendly environment provided by Sketchfab for defining the properties (e.g. opacity, textures, luminosity, etc.) of the objects that make up the model and for the post-processing of the model to improve the rendering and highlight specific features of interest.

In the framework of 3DMAP-VR, we have realized several galleries of models, that can be also explored in VR, on the Sketchfab\footnote{\url{https://sketchfab.com/sorlando}} site and on Artstation\footnote{\url{https://www.artstation.com}}, a leading showcase platform for art and design (see examples in Fig.~\ref{fig:artstation}). Through these galleries, we promote wide dissemination of astrophysical model results for both scientific and public outreach purposes.

\begin{figure*}[t!]
\resizebox{\hsize}{!}{\includegraphics[clip=true]{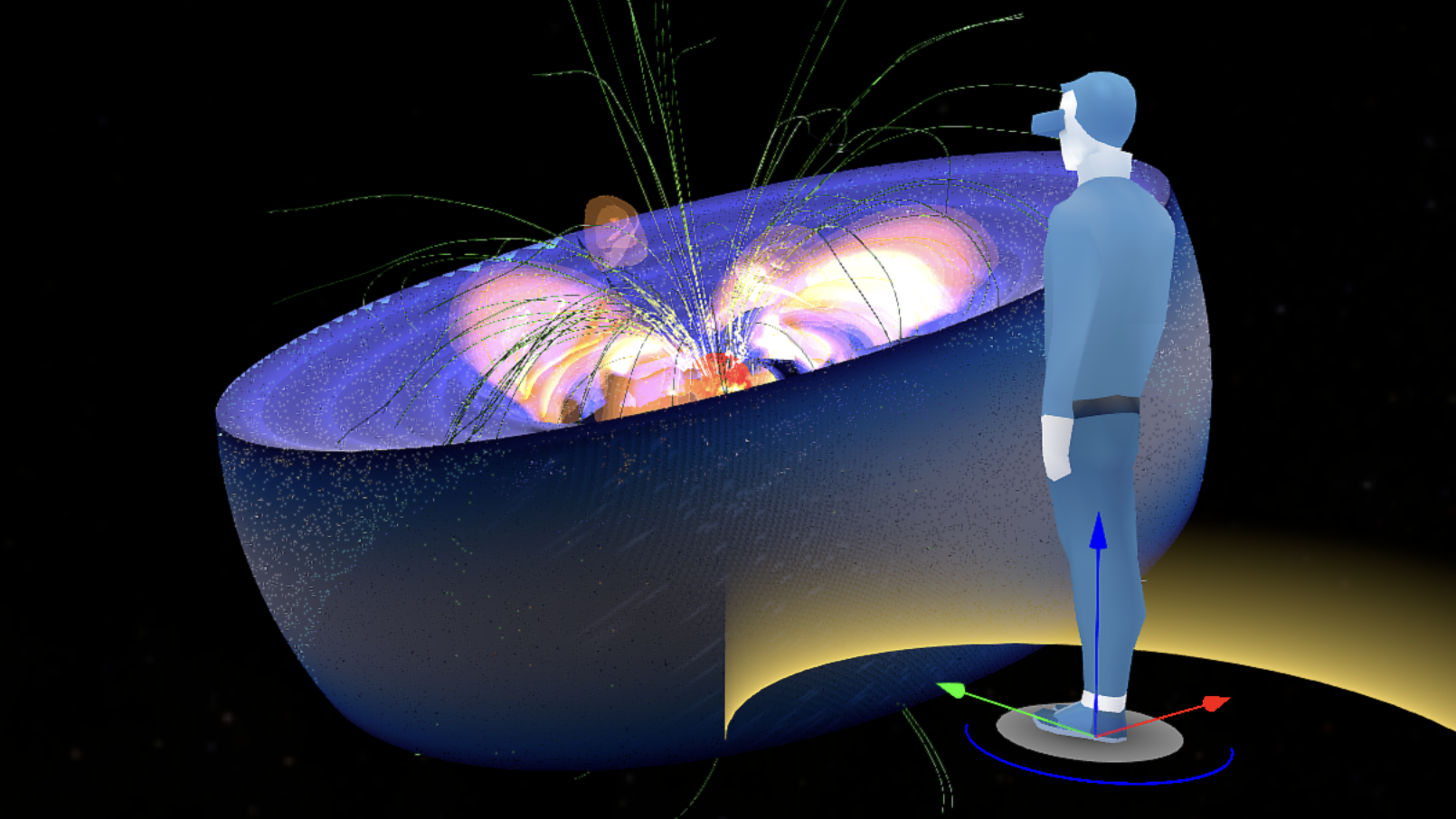}
\includegraphics[clip=true]{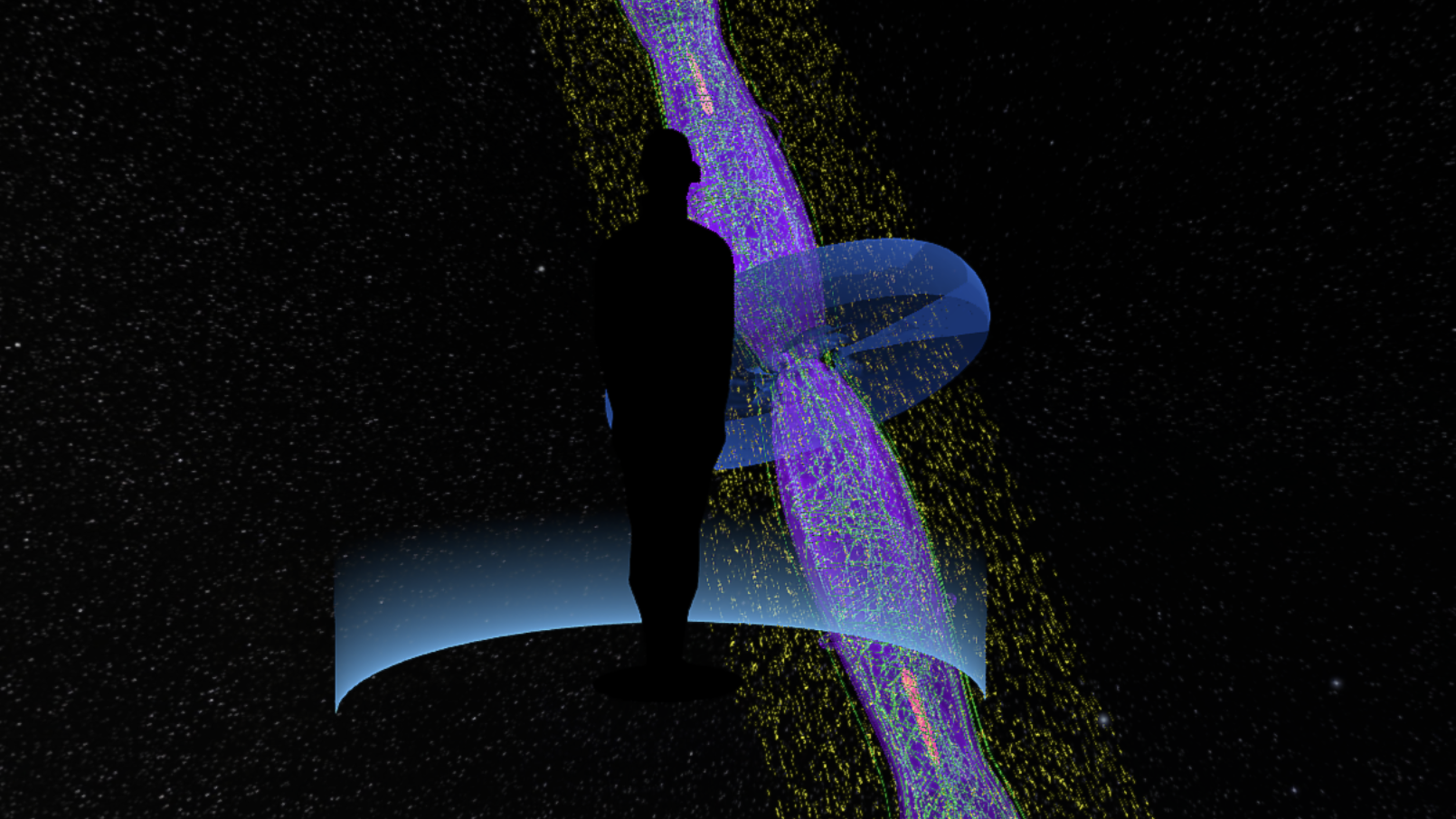}}
\resizebox{\hsize}{!}{\includegraphics[clip=true]{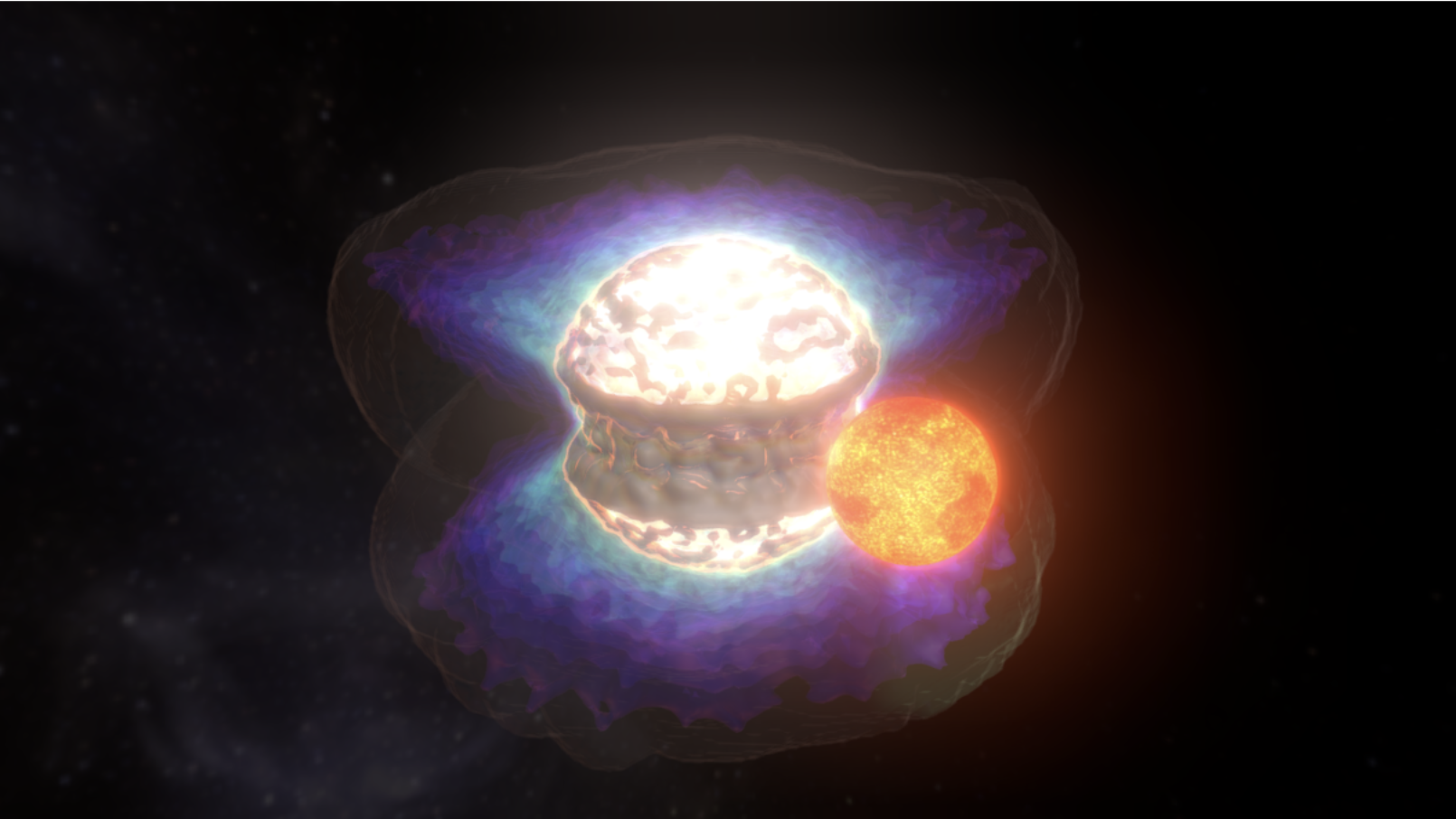}
\includegraphics[clip=true]{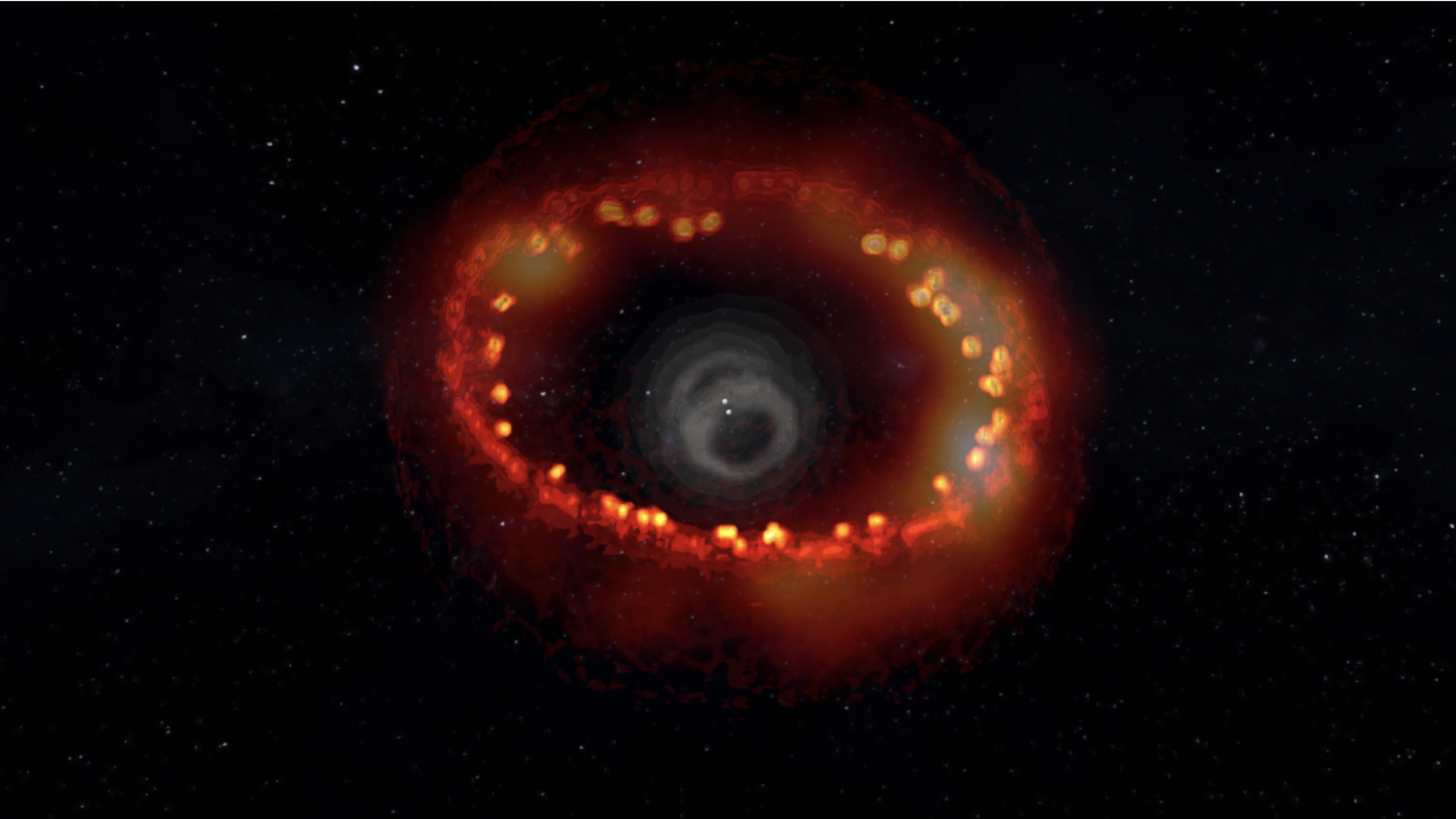}}
\caption{\footnotesize Examples of scientific models available on the Sketchfab platform: a young accreting star (upper left panel; e.g., \citealt{2019A&A...624A..50C}), a protostellar jet (upper right; e.g., \citealt{2016A&A...596A..99U}), the 2010 outburst of nova U Scorpii (lower left; e.g., \citealt{2010ApJ...720L.195D}), and the remnant of SN 1987A (lower right; e.g., \citealt{oon20}). The corresponding interactive graphics can be visited at the following links: \url{https://skfb.ly/6RPnX}, \url{https://skfb.ly/6Rq69}, \url{https://skfb.ly/6Rn8u}, 
(\url{https://skfb.ly/6YNNE}.}
\label{fig:sketch}
\end{figure*}

More specifically, in the Sketchfab platform we realized the gallery ``Universe in hands\footnote{\url{https://skfb.ly/oxHxq}}'' dedicated to 3D HD/MHD models developed by our team for scientific purposes and published in international scientific journals. For instance, thanks to these models, we have investigated (see Fig.~\ref{fig:sketch}): accretion phenomena in young stellar objects (e.g., \citealt{2011MNRAS.415.3380O, 2019A&A...624A..50C}); protostellar jets (e.g., \citealt{2011ApJ...737...54B, 2016A&A...596A..99U}); nova outbursts (e.g., \citealt{2010ApJ...720L.195D, 2012MNRAS.419.2329O}); the outcome of SN explosions (e.g., \citealt{onf20,oon20}).

The assets derived from scientific simulations have been very successful during public outreach events and among non-experts. This positive reception encouraged us to go beyond numerical simulations. So we started creating new models not based on numerical simulations but illustrating astrophysical objects on the base of our current knowledge of these phenomena. The first collection of this new class of resources was ``The art of Astrophysical Phenomena\footnote{\url{https://skfb.ly/oxHx9}}'' where it is possible to visit and explore artists’ views of astrophysical phenomena and objects. Then we created two additional collections: ``Anatomy of Astrophysical Objects\footnote{\url{https://skfb.ly/oxHxz}}'' and ``The Science of Science Fiction\footnote{\url{https://skfb.ly/oxHxv}}''. The first collection reports schematic representations of the structure of astrophysical objects based on our knowledge. The assets of the second collection spot famous science fiction movies to highlight whether and in which parts they get the science right (thus providing accurate and plausible science).

The resources produced in the framework of the 3DMAP-VR project were also used for dissemination purposes by other international scientific institutes. For example, a few assets belonging to the galleries mentioned above were used by NASA to make a series of 3D visualizations of astronomical objects observed with X-ray observatories\footnote{\url{https://youtu.be/wIMB-D9l3I0}} and 3D print kits have been produced for people with visual impairments (e.g., \citealt{2019JCOM...18..201A, 2020FrEaS...8..541A}). Other models illustrating five of the most popular supernova remnants (SNRs) in our Galaxy were used in the standalone application ``StarBlast'' (see Sect.~\ref{sec3}) which exploits the power of VR for the dissemination and educational projects within the scientific activities of the international PHAROS project. The assets were also used to create a series of videos describing astrophysical objects and phenomena, available in Italian ({\em SocialMente: condividiAMO l’universo}\footnote{\url{https://youtube.com/playlist?list=PLITL-h-D-WYQQ7UDMP3CNhp_eYMN94UPU}}) and in English ({\em Universe in hands}\footnote{\url{https://youtube.com/playlist?list=PLITL-h-D-WYRfehOoiTRx_bZ5YFI-bZ8w}}). Finally, other assets have been used to create exhibitions in public outreach events and others have been requested for planetarium exhibitions.

\section{StarBlast: a VR tour of the outcome of stellar explosions}
\label{sec3}

\begin{figure*}[t!]
\resizebox{\hsize}{!}{\includegraphics[clip=true]{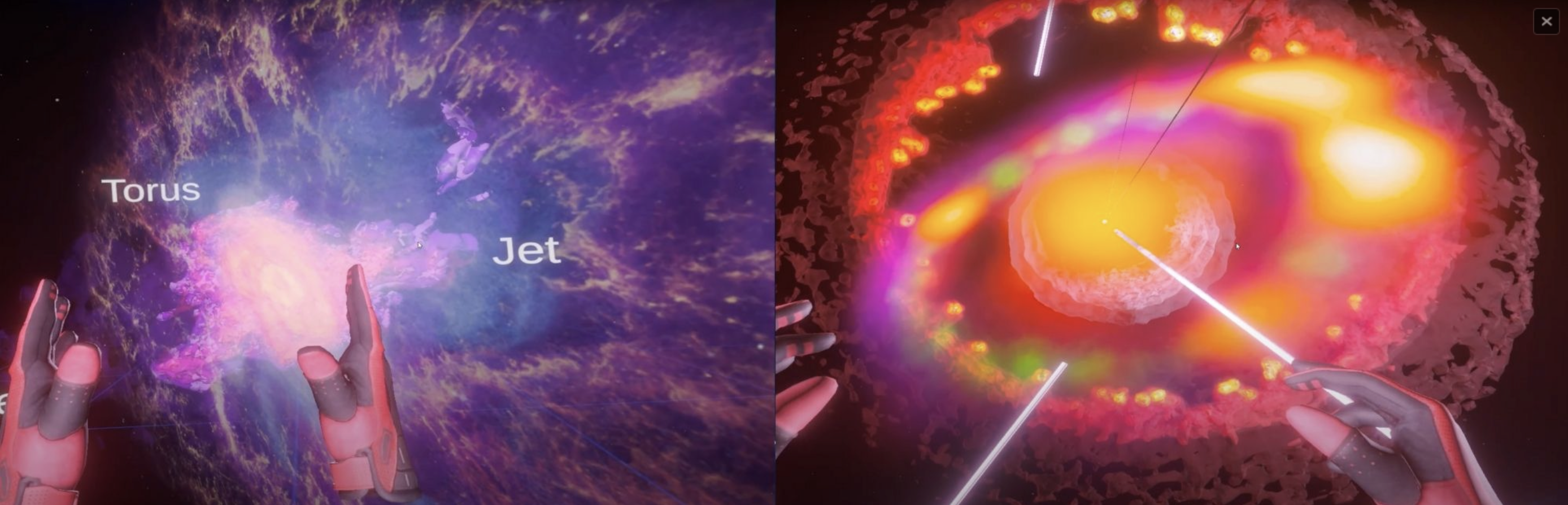}}
\caption{\footnotesize Two examples of scenes in the StarBlast app showing the Crab nebula (on the left; \citealt{oda16}) and SN 1987A (on the right; \citealt{oon20}).}
\label{fig:starblast}
\end{figure*}

As mentioned before, the remnants of SN explosions are characterised by a strong complexity in the distribution of their physical and chemical properties. The adiabatic cooling of the expanding ejecta results in very cold, and tenuous, metal-rich material, which is commonly considered as an important ``dust factory" for our Galaxy. At the same time, the supersonic expansion drives strong (highly supersonic and super-Alfvenic) shocks, which heat the interstellar material and the ejecta themselves up to X-ray emitting temperature. The presence of such a multi-phase material is further complicated by intrinsic anisotropies in the ejecta properties (velocity field, density, chemical composition, pressure, etc) and by the knotty and complex interstellar environment where the remnants evolve (e.g., \citealt{vin20}).

Deciphering the structure of SNRs is crucial for our understanding of the physics governing their formation and evolution. At the same time, the physical modeling of these objects relies on multi-dimensional HD/MHD simulations, whose outputs pose a serious challenge for standard data visualization tools. The intrinsic complexity of the system under exam, indeed requires a novel approach for its proper visualization and analysis. VR provides an immersive experience which enhances the capability of grasping details of the simulations and extract information on SNRs from the numerical models. 

To this end, we developed ``StarBlast: a VR tour of the outcome of stellar explosions" (hereafter StarBlast), a standalone app which exploits the power of VR to offer an immersive experience inside 3D simulations and actual observations of SNRs and pulsar wind nebulae (PWNe). StarBlast took advantage of the successful results obtained within the 3DMAP-VR project (see Sect. \ref{sec2}). The development of the app was supported by the COST Action PHAROS\footnote{\url{http://www.pharos.ice.csic.es/}}, thanks to the proposal ``Real power of virtual reality: developing a standalone app for outreach and teaching activities'' (PI M. Miceli), and by INAF-Osservatorio Astronomico G. S. Vaiana of Palermo and the University of Palermo.

StarBlast was carefully designed to meet the needs of different typologies of user, so as to be adopted for outreach, teaching and research projects. The app is easily accessible to the general public. A voice over (available in English, Italian and Spanish) offers a simple but comprehensive description of the objects during the navigation. Moreover, the very smooth rendering of the astrophysical objects, and the user-friendly control system allow the users to ``naturally" interact with SNRs and PWNe, by literally using the hands (see Fig.~\ref{fig:starblast}).  On the other hand, StarBlast offers also a deeper level of fruition, specifically conceived for students, who can get a clear view of the system and a more intuitive grasp on the physics governing it. We successfully adopted StarBlast in many public outreach events and as a support for lectures in the courses of ``Astrophysics'' and ``Stellar Evolution" at the University of Palermo (Italy). Finally, the app offers a substantial support for researchers, who can access state of the art MHD models and observations (together with the corresponding references, which are provided for all objects), to get a deeper level of diagnostics.

The objects currently included in the app are: SN 1987a (based on the results presented in \citealt{omp16,omp19,mob19}) with its putative PWN (\citealt{gmo21,gmo22}), SN 1006 (\citealt{bom11,obm12,mop16}), the Crab Nebula (\citealt{oda16}), Cassiopeia A (\citealt{omp16,owj21,owj22}) and IC 443 (\citealt{uog21}). However, the library of astrophysical sources in StarBlast can be easily enhanced with new objects.

StarBlast works with the most diffused VR headsets (provided that SteamVR is installed) and is freely available (download link in the footnote\footnote{\url{https://axt.oapa.inaf.it/starblast/}}). 

\section{A VR-based tool for the analysis of numerical scientific simulations}
\label{sec4}

In addition to the immersive visualization of 3D HD/MHD models, VR technologies can be exploited for a deep analysis of these models. Analysis tools based on VR would enable the exploration of the 3D models in an interactive way, achieving a highly detailed analysis that cannot be performed with classical data visualization and analysis platforms. However, VR-based tools specifically tailored for the analysis of numerical simulations are not routinely used in the scientific community and it is necessary to develop ad-hoc techniques and methods.

As a first step, we started exploring software frameworks that could help develop this kind of tools, probing their capabilities in terms of dealing with the data we are interested in and implementing analysis functions. The high-interactivity and the powerful visualization capabilities required, lead us to consider the use of a game engine, a software framework oriented to the development of video-games. Usually, a modern game engine includes features like a level/map editor, a renderer, a physics engine, a collision manager, a scripting system, etc. The use of such a framework allows to accelerate game development, since developers can make use of the mechanics embedded in the engine and focus on the game specific contents and rules. Moreover, engines often provide platform abstraction, meaning that a developed game can seamlessly be deployed on multiple platforms (Windows, Mac, console, smartphone, etc.). Now some game engines have evolved for all-purpose 3D creation, including film-making, architecture, and “serious game” simulations. For our purpose, we selected Unreal Engine (UE; \citealt{unreal_engine}), v.5, for its powerful graphics, for having a powerful scripting language (C++), and for its licensing policy allowing the use of full-featured software for free with no royalty due for developed products that generate a lifetime revenue $<~1~M$. Other works already proposed the use of UE for scientific data visualization and analysis \citep{Friese2008, Marsden2020, Smith2020, unreal_chimera} but, to our knowledge, no VR tool has yet been developed for interactively performing data analysis on models described by data-cubes, such as the results of HD/MHD simulations or similar 3D data (e.g. computer tomography scans).

We tested the feasibility of two operations, that we consider the first steps for developing VR analysis tools: volumetric rendering visualization of our models, and performing interactive data processing leveraging an external analysis software from within the UE generated VR environment.

\begin{figure}[t!]
\resizebox{\hsize}{!}{\includegraphics[clip=true]{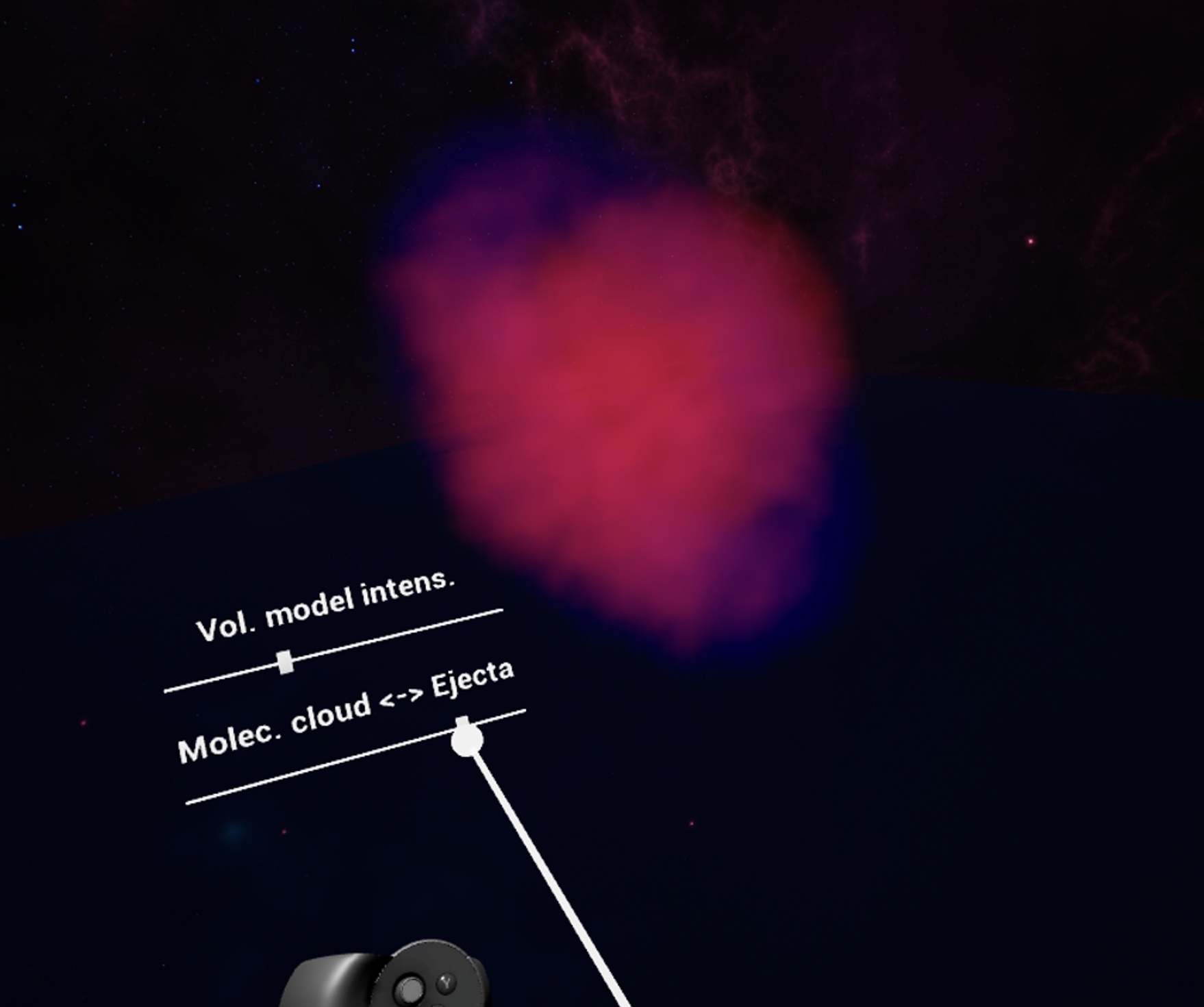}}
\caption{\footnotesize Volumetric rendering of an SNR IC~443 model \citep{uog21} within a UE generated VR environment. The model was produced by a HD simulation and contains two components, the ejecta density on the red channel and the molecular cloud density in the blue channel. The floating interface allows to change in real-time the emissive intensity of the model and the intensity balance between the two components.}
\label{Fig_volumetricRendering}
\end{figure}

Volumetric rendering is computationally expensive, since it implies a full volumetric ray-tracing: the interaction of light with every volume element (voxel) has to be dynamically calculated in real-time in order to reconstruct a correct visual representation. A real-time volumetric rendering of a science simulation output data-cube, in VR, is a task that many 3D visualization software and platforms are not able to perform. UE is highly optimized for rendering, and we verified that it is indeed able to render a VR volumetric model. Moreover, it is possible to change model visualization parameters in real-time, allowing a high level of user interaction. The first step to achieve the volumetric rendering is to use the simulation output data, combining its variables, to generate 3D arrays containing representations of the model (e.g., the total mass density, the mass density of specific components, the temperature, the X-ray emission, etc.). We used a model of the SNR IC~443 \citep{uog21} to create two 3D arrays of $512^3$ elements that represent the mass density of the SN ejecta, and the mass density of an interacting molecular cloud. Each 3D array is then sliced along one dimension obtaining a sequence of tiled slices. A 2D image is generated, containing in its red and blue channels the tiled slices from the two 3D arrays. This procedure, that we perform using a Python script, allows us to convert the visualization data to an image, a format that UE can import as 2D Texture. Since an image has four channels (red, green, blue, transparency), it is possible to embed the data of up to four visual representations. Within UE, we convert the data to a Volumetric Texture, that in turn is used as a base to create a Volumetric Material. A Material, assigned to an object within a virtual environment, describes how it will be displayed, defining its color, transparency, emissivity, etc. Assigning the Volumetric Material to a base object placed in the VR environment, in our case a sphere, allows UE to perform the volumetric rendering of the 3D model, as shown in Fig.~\ref{Fig_volumetricRendering}. We also implemented a control panel, visible in the virtual environment as floating over the user wrist, featuring sliders that can control in real-time some properties of the rendered model. We tested, as a proof-of-concept, the variation of the emission intensity of the model and the intensity balance between the two components of the model (see Fig.~\ref{Fig_volumetricRendering}). The result is very promising, indicating that it is possible to modify visualizations properties on the fly while keeping a fluid rendering in VR. We also implemented the possibility to interact with the model, grabbing it and moving and rotating it as if it were held in the user hand, and to enlarge it or scale it down using intuitive hand gestures.

\begin{figure}[t!]
\resizebox{\hsize}{!}{\includegraphics[clip=true]{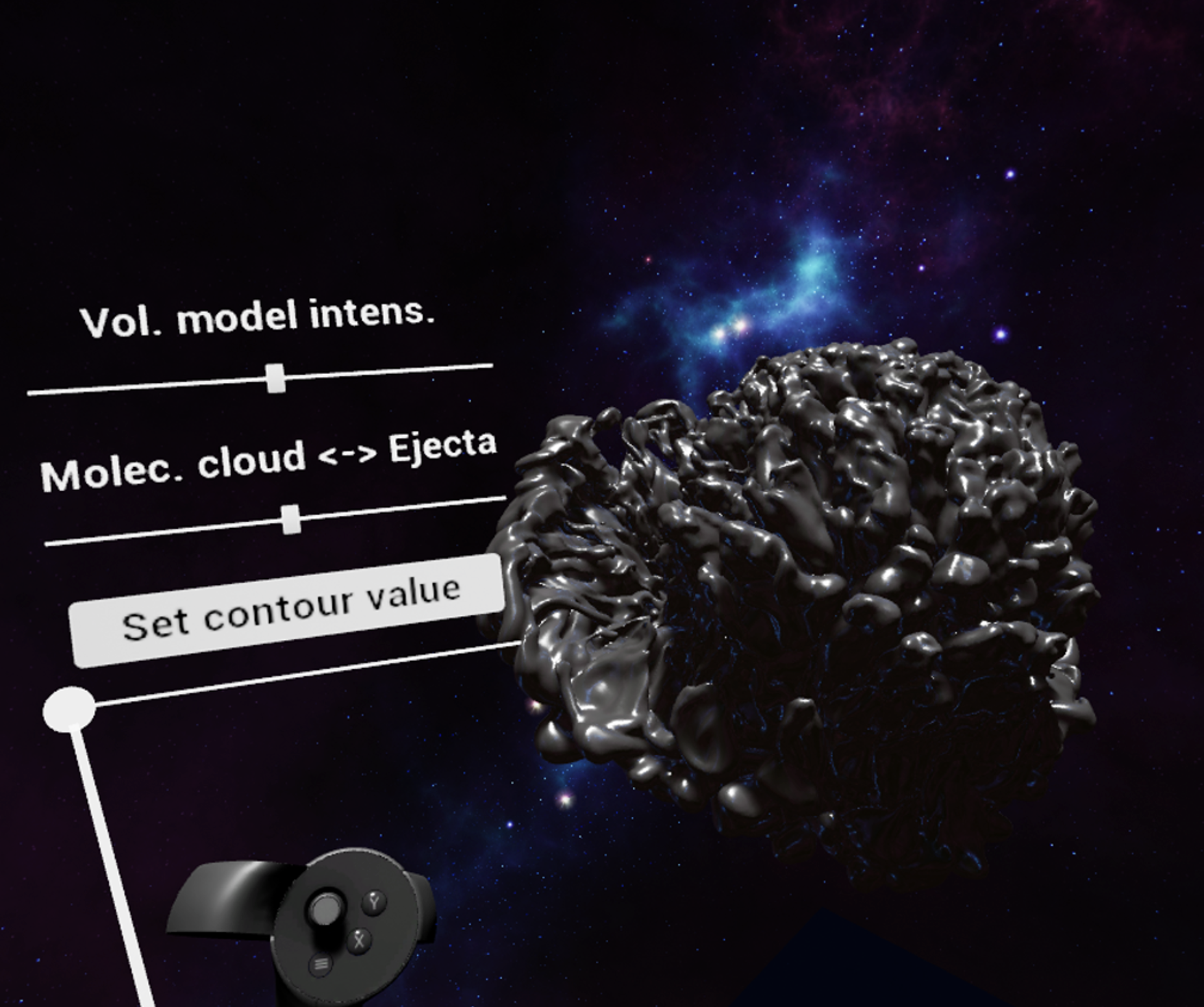}}
\caption{\footnotesize Iso-density contour of SNR IC~443 ejecta \citep[see][]{uog21} generated with ParaView and rendered in VR within UE. The floating interface allows to change the density value of the contour. The "Set contour value" button executes an analysis pipeline, leveraging ParaView to create an updated contour that is then rendered in the environment.}
\label{Fig_ParaviewContour}
\end{figure}

As a second feasibility test we checked if it is possible to use UE in synergy with an external data analysis software. Building a complete analysis suite fully within UE, while technically feasible, would require a huge effort, a lot of manpower with high programming skills and deep understanding of data manipulation algorithms and methods. We investigated the use of ParaView \citep{Paraview}, widely used advanced software for analysis of 3D data-cubes, from within UE. ParaView exposes C APIs that could be exploited from UE for direct integration. For a preliminary concept demonstrator, however, we adopted an indirect and simpler approach.  From the VR environment, built with UE, we seamlessly call a Python script that uses ParaView to load a 3D model (HD/MHD simulation output), applies an analysis pipeline (i.e. extracts a iso-density surface, a contour), and save the resulting 3D object to the disk. The newly created 3D asset is then loaded and rendered within the VR environment (see the contour of SNR IC~443 in Fig.~\ref{Fig_ParaviewContour}). We can pass parameters to the script to modify the analysis pipeline, e.g. changing the density used to extract the iso-surface using the bottom slider shown in Fig.~\ref{Fig_ParaviewContour}. The delay between the trigger in the VR world and the visualization of the model resulting from the analysis depends on the size of the data-cube, on the computer performances, and on the analysis pipeline; for extracting a contour from a $256^3$ elements data-cube we found a delay of about 4 s.

We thus demonstrated the feasibility of two fundamental tasks with UE, the volumetric rendering of a model created for scientific purposes, and interacting with an external analysis software to perform data processing. The next steps for developing VR analysis tools are the C++ implementation, within UE, of some analysis operations (e.g. slicing), and the integration of ParaView, through its APIs, to sensibly speed up the analysis process.

\section{Conclusions}

In this contribution we reported the recent achievements obtained at INAF-Osservatorio Astronomico di Palermo in the development of VR-based tools aimed at the analysis and visualization of 3D scientific models describing astronomical objects and phenomena. These tools allow to explore the models interactively in an immersive fashion, resulting in highly detailed analysis that cannot be performed with traditional data visualization (e.g., based on screen displays) and analysis platforms. In addition our tools have been proved to be very powerful in the production of high-impact VR content for efficient audience engagement and outreach.

\begin{acknowledgements}
We thank the anonymous referee for the careful reading of the paper. S.O. and M.M. acknowledge financial contribution from the PRIN INAF 2019 grant ``From massive stars to supernovae and supernova remnants: driving mass, energy and cosmic rays in our Galaxy'' and the INAF mainstream program ``Understanding particle acceleration in galactic sources in the CTA era''. The development of the app StarBlast was supported by the COST Action PHAROS, thanks to the proposal ``Real power of virtual reality: developing a standalone app for outreach and teaching activities''. We acknowledge the high performance computing (HPC) facility at CINECA through the ISCRA programme and the SCAN (Sistema di Calcolo per l'Astrofisica Numerica) HPC facility at INAF-Osservatorio Astronomico di Palermo for the availability of HPC resources and support.
\end{acknowledgements}

\bibliographystyle{aa}
\bibliography{references}

\begin{thebibliography}{32}
\expandafter\ifx\csname natexlab\endcsname\relax\def\natexlab#1{#1}\fi

\bibitem[{{Arcand} {et~al.}(2019){Arcand}, {Jubett}, {Watzke}, {Price},
  {Williamson}, \& {Edmonds}}]{2019JCOM...18..201A}
{Arcand}, K., {Jubett}, A., {Watzke}, M., {et~al.} 2019, JCOM Journal of
  Science Communication, 18, 18040201

\bibitem[{{Arcand} {et~al.}(2020){Arcand}, {Price}, \&
  {Watzke}}]{2020FrEaS...8..541A}
{Arcand}, K.~K., {Price}, S.~R., \& {Watzke}, M. 2020, Frontiers in Earth
  Science, 8, 541

\bibitem[{{Bocchino} {et~al.}(2011){Bocchino}, {Orlando}, {Miceli}, \&
  {Petruk}}]{bom11}
{Bocchino}, F., {Orlando}, S., {Miceli}, M., \& {Petruk}, O. 2011, \aap, 531,
  A129

\bibitem[{{Bonito} {et~al.}(2011){Bonito}, {Orlando}, {Miceli}, {Peres},
  {Micela}, \& {Favata}}]{2011ApJ...737...54B}
{Bonito}, R., {Orlando}, S., {Miceli}, M., {et~al.} 2011, \apj, 737, 54

\bibitem[{{Colombo} {et~al.}(2019){Colombo}, {Orlando}, {Peres}, {Reale},
  {Argiroffi}, {Bonito}, {Ibgui}, \& {Stehl{\'e}}}]{2019A&A...624A..50C}
{Colombo}, S., {Orlando}, S., {Peres}, G., {et~al.} 2019, \aap, 624, A50

\bibitem[{{Drake} \& {Orlando}(2010)}]{2010ApJ...720L.195D}
{Drake}, J.~J. \& {Orlando}, S. 2010, \apjl, 720, L195

\bibitem[{Epic~Games(2023)}]{unreal_engine}
Epic~Games, I. 2023, Unreal Engine

\bibitem[{Friese {et~al.}(2008)Friese, Herrlich, \& Wolter}]{Friese2008}
Friese, K.-I., Herrlich, M., \& Wolter, F.-E. 2008, in New Frontiers for
  Entertainment Computing, ed. P.~Ciancarini, R.~Nakatsu, M.~Rauterberg, \&
  M.~Roccetti (Boston, MA: Springer US), 11--22

\bibitem[{{Fryxell} {et~al.}(2000){Fryxell}, {Olson}, {Ricker}, {Timmes},
  {Zingale}, {Lamb}, {MacNeice}, {Rosner}, {Truran}, \&
  {Tufo}}]{2000ApJS..131..273F}
{Fryxell}, B., {Olson}, K., {Ricker}, P., {et~al.} 2000, \apjs, 131, 273

\bibitem[{{Greco} {et~al.}(2021){Greco}, {Miceli}, {Orlando}, {Olmi},
  {Bocchino}, {Nagataki}, {Ono}, {Dohi}, \& {Peres}}]{gmo21}
{Greco}, E., {Miceli}, M., {Orlando}, S., {et~al.} 2021, \apjl, 908, L45

\bibitem[{{Greco} {et~al.}(2022){Greco}, {Miceli}, {Orlando}, {Olmi},
  {Bocchino}, {Nagataki}, {Sun}, {Vink}, {Sapienza}, {Ono}, {Dohi}, \&
  {Peres}}]{gmo22}
{Greco}, E., {Miceli}, M., {Orlando}, S., {et~al.} 2022, \apj, 931, 132

\bibitem[{Kitware(2022)}]{Paraview}
Kitware, I. 2022, ParaView – Open-source, multi-platform data analysis and
  visualization application

\bibitem[{Marsden \& Shankar(2020)}]{Marsden2020}
Marsden, C. \& Shankar, F. 2020, Universe, 6, 168

\bibitem[{Mayerich {et~al.}(2020)Mayerich, Chen, \& Childs}]{unreal_chimera}
Mayerich, D., Chen, G., \& Childs, H. 2020, Unreal Chimera: An Unreal Engine
  Platform for Scientific Visualization of Next-Generation Data

\bibitem[{{Miceli} {et~al.}(2019){Miceli}, {Orlando}, {Burrows}, {Frank},
  {Argiroffi}, {Reale}, {Peres}, {Petruk}, \& {Bocchino}}]{mob19}
{Miceli}, M., {Orlando}, S., {Burrows}, D.~N., {et~al.} 2019, Nature Astronomy,
  3, 236

\bibitem[{{Miceli} {et~al.}(2016){Miceli}, {Orlando}, {Pereira}, {Acero},
  {Katsuda}, {Decourchelle}, {Winkler}, {Bonito}, {Reale}, {Peres}, {Li}, \&
  {Dubner}}]{mop16}
{Miceli}, M., {Orlando}, S., {Pereira}, V., {et~al.} 2016, \aap, 593, A26

\bibitem[{{Mignone} {et~al.}(2012){Mignone}, {Zanni}, {Tzeferacos}, {van
  Straalen}, {Colella}, \& {Bodo}}]{2012ApJS..198....7M}
{Mignone}, A., {Zanni}, C., {Tzeferacos}, P., {et~al.} 2012, \apjs, 198, 7

\bibitem[{{Olmi} {et~al.}(2016){Olmi}, {Del Zanna}, {Amato}, {Bucciantini}, \&
  {Mignone}}]{oda16}
{Olmi}, B., {Del Zanna}, L., {Amato}, E., {Bucciantini}, N., \& {Mignone}, A.
  2016, Journal of Plasma Physics, 82, 635820601

\bibitem[{{Ono} {et~al.}(2020){Ono}, {Nagataki}, {Ferrand}, {Takahashi},
  {Umeda}, {Yoshida}, {Orlando}, \& {Miceli}}]{onf20}
{Ono}, M., {Nagataki}, S., {Ferrand}, G., {et~al.} 2020, \apj, 888, 111

\bibitem[{{Orlando} {et~al.}(2012){Orlando}, {Bocchino}, {Miceli}, {Petruk}, \&
  {Pumo}}]{obm12}
{Orlando}, S., {Bocchino}, F., {Miceli}, M., {Petruk}, O., \& {Pumo}, M.~L.
  2012, \apj, 749, 156

\bibitem[{{Orlando} \& {Drake}(2012)}]{2012MNRAS.419.2329O}
{Orlando}, S. \& {Drake}, J.~J. 2012, \mnras, 419, 2329

\bibitem[{{Orlando} {et~al.}(2019{\natexlab{a}}){Orlando}, {Miceli}, {Petruk},
  {Ono}, {Nagataki}, {Aloy}, {Mimica}, {Lee}, {Bocchino}, {Peres}, \&
  {Guarrasi}}]{omp19}
{Orlando}, S., {Miceli}, M., {Petruk}, O., {et~al.} 2019{\natexlab{a}}, \aap,
  622, A73

\bibitem[{{Orlando} {et~al.}(2016){Orlando}, {Miceli}, {Pumo}, \&
  {Bocchino}}]{omp16}
{Orlando}, S., {Miceli}, M., {Pumo}, M.~L., \& {Bocchino}, F. 2016, \apj, 822,
  22

\bibitem[{{Orlando} {et~al.}(2020){Orlando}, {Ono}, {Nagataki}, {Miceli},
  {Umeda}, {Ferrand}, {Bocchino}, {Petruk}, {Peres}, {Takahashi}, \&
  {Yoshida}}]{oon20}
{Orlando}, S., {Ono}, M., {Nagataki}, S., {et~al.} 2020, \aap, 636, A22

\bibitem[{{Orlando} {et~al.}(2019{\natexlab{b}}){Orlando}, {Pillitteri},
  {Bocchino}, {Daricello}, \& {Leonardi}}]{2019RNAAS...3..176O}
{Orlando}, S., {Pillitteri}, I., {Bocchino}, F., {Daricello}, L., \&
  {Leonardi}, L. 2019{\natexlab{b}}, Research Notes of the American
  Astronomical Society, 3, 176

\bibitem[{{Orlando} {et~al.}(2011){Orlando}, {Reale}, {Peres}, \&
  {Mignone}}]{2011MNRAS.415.3380O}
{Orlando}, S., {Reale}, F., {Peres}, G., \& {Mignone}, A. 2011, \mnras, 415,
  3380

\bibitem[{{Orlando} {et~al.}(2022){Orlando}, {Wongwathanarat}, {Janka},
  {Miceli}, {Nagataki}, {Ono}, {Bocchino}, {Vink}, {Milisavljevic}, {Patnaude},
  \& {Peres}}]{owj22}
{Orlando}, S., {Wongwathanarat}, A., {Janka}, H.~T., {et~al.} 2022, \aap, 666,
  A2

\bibitem[{{Orlando} {et~al.}(2021){Orlando}, {Wongwathanarat}, {Janka},
  {Miceli}, {Ono}, {Nagataki}, {Bocchino}, \& {Peres}}]{owj21}
{Orlando}, S., {Wongwathanarat}, A., {Janka}, H.~T., {et~al.} 2021, \aap, 645,
  A66

\bibitem[{Smith \& Greenwood(2020)}]{Smith2020}
Smith, M. \& Greenwood, M. 2020, in Transactions of the American Nuclear
  Society - Volume 123 ({AMNS})

\bibitem[{{Ustamujic} {et~al.}(2016){Ustamujic}, {Orlando}, {Bonito}, {Miceli},
  {G{\'o}mez de Castro}, \& {L{\'o}pez-Santiago}}]{2016A&A...596A..99U}
{Ustamujic}, S., {Orlando}, S., {Bonito}, R., {et~al.} 2016, \aap, 596, A99

\bibitem[{{Ustamujic} {et~al.}(2021){Ustamujic}, {Orlando}, {Greco}, {Miceli},
  {Bocchino}, {Tutone}, \& {Peres}}]{uog21}
{Ustamujic}, S., {Orlando}, S., {Greco}, E., {et~al.} 2021, \aap, 649, A14

\bibitem[{{Vink}(2020)}]{vin20}
{Vink}, J. 2020, {Physics and Evolution of Supernova Remnants}

\end{thebibliography}

\end{document}